\documentclass[aps,twocolumn,showpacs]{revtex4}

\usepackage{graphicx}%
\usepackage{dcolumn,float}
\usepackage{amsmath,amssymb}

\begin{document}

\title{Single-Block Renormalization Group:  
\\  Quantum Mechanical Problems }

\date{\today}

\author{  
M.A. Mart\'{\i}n-Delgado$^{\star}$, J. Rodriguez-Laguna$^{\ast}$ and
G. Sierra$^{\ast}$
 } 
\address{ 
$^{\star}$Departamento de
F\'{\i}sica Te\'orica I, Universidad Complutense. 28040 Madrid, Spain.
\\ 
$^{\ast}$Instituto de Matem\'aticas y F\'{\i}sica Fundamental, C.S.I.C.,
Madrid, Spain. }

\begin{abstract}

We  reformulate  the density 
matrix renormalization group method (DMRG) 
in terms of  a single block, 
instead of the standard 
left and right blocks used in the construction 
of the superblock. 
This version of the DMRG, which we call the puncture renormalization
group (PRG), makes easy and natural the extension of   
the DMRG to higher dimensional lattices. 
To test numerically this proposal,
we study several quantum 
mechanical models in one, two and three dimensions.
In 1D the performance of 
the standard DMRG is much better than its PRG version,
however for 2D models  
the PRG is more efficient than the DMRG
in a variety of circumstances. 
In 3D the PRG performs also quite well.

\end{abstract}

\pacs{75.10.-b, 05.50.+q, 03.65.-w}% PACS, the Physics and Astronomy Classification Scheme.

\maketitle

\section{Introduction}
\label{sec1:level1}

The elucidation of the low energy  properties in  quantum two-dimensional
strongly correlated systems such as Heisenberg, t-J and Hubbard models
is known to be one of the central issues in modern condensed matter Physics,
and a variety of approximate methods have been devised to treat this type
of problems, not always under a controllable way.
It is believed that by solving these issues, what is at stake is the solution
of the long sought-after mechanism underlying high-$T_c$ superconductivity
in layered cuprate compounds,
just to mention one of the many possible applications.

After several years of improvements and extensions since its
formulation \cite{white1}, \cite{white2}, \cite{noackwhite},
the Density Matrix Renormalization Group (DMRG)
method has become by now a standard numerical tool
in the study of strongly correlated systems.
We shall not dwell upon its defining features here, although
we will do it in next sections when dealing with specific examples.
The reader is referred  to several reviews in the literature, such as
\cite{dresden98}, \cite{whiterep},\cite{karen},
\cite{jaitisi}, \cite{elescorial}.

The method is specially well-suited when the problem under study
is purely one-dimensional or quasi-one dimensional (such as
ladders \cite{dmrgladder}), although applications to small clusters of
quantum two-dimensional lattices have also  been carried out with success
\cite{fermion2Da}, \cite{fermion2Db}, \cite{whitecavo},
\cite{white-scalapino2Da},\cite{white-scalapino2Db},\cite{white-scalapino2Dc}.

Nevertheless, a formulation of the DMRG method in two dimensions
on equal footing as it is conceived in the one-dimensional case, is still
lacking. The performance of the method is  by far more powerful
when the model is formulated in a chain rather than in a two-dimensional
lattice \cite{pang}. There must be deeply rooted  reasons in the setup of the
method to account for this behaviour.

There have been various attempts to address this kind of
higher dimensional extensions using several techniques.
Here we briefly enumerate some of them:
i) the original ``zipping" method for
the DMRG in 2D as introduced in \cite{fermion2Da}. This is
considered the standard method.
It amounts to introduce a one-dimensional sublattice inside the
original 2D lattice in order to perform the sweeping process
characteristic of the finite-system method.
We shall review it in Sect.~\ref{sec2:level1};
ii) the extensions based on the classical Statistical Mechanics
applications of the DMRG on 2D classical lattices \cite{nishino1}
\cite{okunishi},
\cite{nishino2}.
These are based on the Corner Transfer Method.
Recent extensions from 2D to 3D lattices have been
proposed in \cite{nishino3,nishino4}; iii) extensions inspired in the
Matrix Product formulation of the DMRG \cite{ostlund-romer} have
been considered in \cite{rva-dresden}; iv) extensions inspired in
the superblock method \cite{cbrg},\cite{noackwhite}.

Despite  all of these proposals,
we believe that there is still room for improvements regarding
this issue. Specifically, we want to formulate the DMRG in such
a way that going  from one to two, or higher, dimensions is
a natural step. We mean by this that all the components
entering the DMRG need not be modified accordingly  with the dimensionaliy
of the lattice. Quite on the contrary, we pursue a formulation in which
all the elements of the method are independent of the dimensionality.
In other words, we want a dimension-independent 
formulation of the DMRG method.
Moreover, we also envisage the possibility of having the errors arising
from the truncations of the Hilbert space of states kept under control
as the size of the lattice increases, at least to the extend that the
errors are also isotropically distributed.

In this work we have fully developed this desiderata
for quantum mechanical problems in one, two and three dimensions.
Quantum mechanical problems have played a paramount role in the
development of the DMRG \cite{noackwhite,white1}. It is well-known
by now that it was the analysis of the Wilson's RG failure for the
simple particle-in-a-box problem which led to the invention of the
Density Matrix RG. Here we want to follow the same approach and
we present a new formulation of the DMRG in several dimensions
and check its validity with one-particle quantum problems of
several types.

The rationale behind this approach is that if we find  a new
version of the DMRG for 2D quantum lattice problems that fails to reproduce
the low energy physics of the  Quantum Mechanics in 2D,
then that version of the DMRG is doom to failure when applied
to  more complicated quantum many-body problems.

Moreover, as a spinoff of our work, we have devised a numerical method
to solve the Schr\"odinger equation in several dimensions with an
accuracy and efficiency bigger than other numerical methods known so far
such as exact diagonalization techniques. Namely, we can reach lattice
sizes which are out of reach for exact diagonalization, while keeping
the same degree of accuracy.

This paper is  organized as follows: in Sect.~\ref{sec2:level1}
we review the DMRG formulation for quantum mechanical problems
with an arbitrary number of low-energy states kept during the
truncation process.
Then, a new formulation using only one block is presented  and
we compare both formulations by stressing their similarities and
main differences. In Sect.~\ref{sec3:level1} we present an 
account of numerical results to test the performance of the
new RG method under a variety of  circunstances: lattices of
several dimensions, discrete Hamiltonians of different types,
variation in the number of targeted states, etc. 
Section Sec.~\ref{sec4:level1} is devoted to conclusions and further prospects.

%%%%%%%%%%%%%%%%%%%%%%%%%%%%%%%%%%%%%%%%%%%%%%%%%%%%%%%%%%%%%%%%%%

\section{One Block vs. Two Block Formulations of DMRG in Quantum Mechanics}
\label{sec2:level1}

There are several ingredients entering the standard formulation of the
DMRG method \cite{white1}:
the left ($B_L$) and right ($B_R$) blocks which describe the degrees of
freedom of the system and the environment, respectively;
the single sites, one $\bullet$ or two $\bullet \bullet$, connecting the
left and right blocks  which serve as a kind of probes to test the
reaction of the system degrees of freedom to the coupling to the rest
of the environment;
the universe ${\cal U}$,
also called superblock (SB), which contains the description of the effective
degrees of freedom of the whole combined system of blocks and sites,
${\cal U} = B_L \bullet  B_R$ or ${\cal U} = B_L \bullet \bullet B_R$,
at a given step of the RG process;
the ground state (GS) wave function obtained after diagonalizing the
superblock Hamiltonian of the system $H_{SB}$ and which is the so called
target state;
the density matrix $\rho_S$ of the system obtained after projecting the
target state down onto the Hilbert space of the system states;
the sweeping process which consists in moving the probe sites $\bullet \bullet$
from left to right and viceversa, back and forth through the superblock,
and it is responsible for achieving the convergence of the targeted properties
of the system, or in the RG language, for reaching the fixed point structure
of the DMRG transformation.

However, if we want to make a higher dimensional extension
of the method, we need to abstract its most important and relevant pieces
from the rest.
We can do this by setting up the following question:
What is the distinctive feature which makes DMRG essentially different
from the Wilson RG \cite{wilson}
and makes it superior and so successful?

\noindent
The answer is {\it correlation}. It is the introduction of correlations
between blocks by means of the superblock construction
and its constant update through the RG iterations 
by means of the sweeping process, that allows the DMRG 
to essentially capture the strong quantum fluctuations
present in low dimensional systems.

\noindent Then, a second question arises: What is the role played by the
density matrix? In a one-dimensional world such as a spin chain, one
probing site $\bullet$ naturally divides the chain into two halves
according to the scheme ${\cal U} = B_L \bullet  B_R$. The correlation
between blocks is restored by means of the density matrix constructed
out of the GS of the universe  (superblock). What is important here
is the restoration of correlations between blocks, no matter whether
it is achieved with the density matrix or with another type of construct.

When we want to extend this construction to higher dimensions we
inmediately run into some problems. For example,
in a two-dimensional world such as a
quantum spin model on the plane, a single probing point $\bullet$
no longer divides the world into two halves. If we want to recover
this splitting we naturally need to substitute the point by a line
$|$, in such a way as to divide the universe with the following
scheme ${\cal U} = B_L | B_R$.
Thus, in 1D a probing point is needed to do the job while in 2D
it is  a probing line.

\noindent We observe  a big difference between 1D and 2D,  
as far as the DMRG splitting is concerned:
a probing point is something manageable for it has a reduced number of
degrees of freedom while a probing line is not directly tractable for
it contains a huge set of states. Actually, a line poses a non-trivial
quantum problem by itself. Thus, it seems that we need to stick somehow
to the use of probing sites instead of lines
even if we want to make higher dimensional extensions for the DMRG.

From this preliminary discussion,
we arrive at the following picture for the basic operations
in the DMRG method:

\noindent {\bf Cutting Process}: this amounts to splitting the lattice into
probing sites and blocks so that the superblock is made up of these
two type of components.

\noindent {\bf Sweeping Process}: this amounts to moving the probing sites
throughout the superblock at each step of the RG process and
updating the content of the blocks in the cutting process.

\noindent It is quite apparent that there is a lot of room to implement these
ingredients. However, we shall present here one of such a schemes and we shall
do it by comparing it with the standard approach of  the DMRG in 2D.
The new scheme will be based on a simpler superblock structure made up of
only one block and only one probing site, namely ${\cal U}=B\bullet$.
We shall refer to this site  as a
{\it puncture} to distinguish it from the usual probing points in the DMRG.
Thus, we shall also refer to this kind of single-block DMRG method as
a Punctured Renormalization Group (PRG) method.
This situation is in sharp contrast with the standard DMRG program which always
employs two blocks to build the superblock.
Later we shall see that the number of punctures can be more than one.
The primary idea behind the PRG version is that the block $B$ will give
an effective description of all the low energy degrees of freedom
except for those associated to the puncture $\bullet$.

%%%%%%%%%%%%%%%%%%%%%%%%%%

\noindent
Specifically, the problem we want to solve using DMRG techniques is the
following Schr\"odinger equation for one single particle in several dimensions:

\begin{equation}
H \Psi =  E \Psi,
\label{1}
\end{equation}

\noindent with the quantum lattice Hamiltonian given by,

\begin{equation}
H_{{\bf n},{\bf m}} =
\begin{cases}
2/h^2 + V_{{\bf n}} & \quad {\bf n}={\bf m} \\
-1/h^2              & \quad \parallel {\bf n}-{\bf m} \parallel  = 1 \\
0                   & \quad \text{otherwise};
\end{cases}
\label{2}
\end{equation}

\noindent where ${\bf n}, {\bf m}$ are vectors of integer
components in a square lattice of dimension $d=1,2,3$; $V_{\bf
n}$ is the local potential at site ${\bf n}$, $h$ is the lattice
spacing and $\parallel . \parallel$ denotes the Euclidean norm
and we select the nearest-neighbour points. This corresponds to a
well-known discretization procedure of the Schr\"odinger equation
defined in the continuum, $H \Psi = (-\Delta_{{\bf x}} + V({\bf
x})) \Psi({\bf x}) = E \Psi$, with ${\bf x} = h {\bf n}$ and each
of the components ranging as ${\bf n}_i = 1,2,\ldots,N_i$ $\forall
i=1,\ldots,d$, where $N_i$ is the number of sites in each direction
of the square lattice. The size $L_i$ in each direction is $L_i = x_N
- x_1 = h (N_i-1)$. Let $N$ be the total number lattice sites,
i.e., $N=\prod_{i=1}^d N_i$. The continuum limit is recovered as
the double limit $h \rightarrow 0$, $N_i \rightarrow \infty$
leaving fixed $L_i=N_i \times h$. We shall not be
interested here in this limit \cite{qm-dmrg} and thus we set $h=1$.

Let us next review the finite-system DMRG algorithm which is used to
obtain a reduced set of low energy states for the Schr\"odinger Hamiltonian
defined in (\ref{1}) \cite{qm-dmrg}, \cite{delta}, \cite{whitebook}.
For the sake of concreteness, let us assume that our lattice is
two-dimensional. The one- and three-dimensional cases will appear as a
restriction and an extrapolation, respectively, from this 2D case, 
as far as the DMRG implementation is concerned.

In the following we shall introduce the different components of the
PRG method after having reviewed the corresponding analogous components
of the standard DMRG program. In this fashion we shall emphasize the
main similarities and differences between both formulations.

\subsection{ Superblock Decomposition of the Lattice}

\begin{figure}
\centering
\includegraphics[width=5.5 cm]{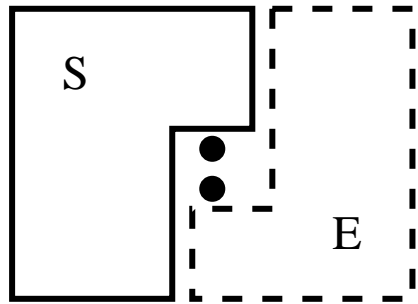}% Here is how to import EPS art
\caption{ Superblock decomposition of a 2D lattice in the standard DMRG
method. The two central black circles represent the probing sites according
to the scheme ${\cal U} = B^L \bullet \bullet B^R$
and S stands for the system block 
and E for the environment block.}
\label{fig1}
\end{figure}

The most common DMRG-decomposition of the lattice is that of a
superblock formed by two blocks and two probing sites:
${\cal U} = B_{l}^L \bullet \bullet B_{N-l-2}^R$
where the subscripts $l$ and $N-l-2$ denote the number of sites
inside each left and right blocks such that
their sum - plus two -  equals the total number of lattice sites $N$.
The index $l$ denotes the iteration step of the RG process.
In Fig.~\ref{fig1} a typical 
superblock decomposition associated to the standard
DMRG is depicted in two dimensions.

%%%%%%%%%%%%%%%%%%%%%%%%%%%%%%%%%%%%%%%%%%%%%%%%%%%%%%%%%%%%%%%%%%%%%%%%%%

%%%%%%%%%%%%%%%%%%%%%%%%%%%%%%%%%%%%%%%%%%%%%%%%%%%%%%%%%%%%%%%%%%%%%%%%%%%%%

%%%%%%%%%%%%%%%%%%%%%%%%%%%%%%%%%%%%%%%%%%%%%%%%%%%%%%%%%%%%%%%%%%%%%%%%%%

\begin{figure}
\centering
\includegraphics[height=5 cm]{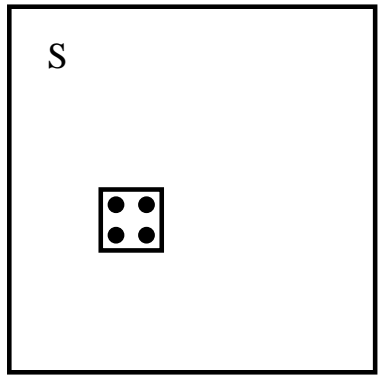}% Here is how to import EPS art
\caption{ Superblock decomposition of a 2D lattice for the PRG
method. The  black circles represent the puntures, which in this
case are four, and S stands for the system block.}
\label{fig2}
\end{figure}

%%%%%%%%%%%%%%%%%%%%%%%%%%%%%%%%%%%%%%%%%%%%%%%%%%%%%%%%%%%%%%%%%%%%%%%%%%%%%

In the PRG version the superblock decomposition will be of the
form ${\cal U} = B_l \bullet$, where $B_l$ stands for a block containing
$N-1$ sites at the step $l$ of the RG process and the site
will be also denoted as $\bullet_l$. Hence the size 
of the block $B_l$ remains unchanged as we vary $l$.
 
In Fig.~\ref{fig2} it is shown one of 
these superblock decompositions for a case
with 4 punctures.

\noindent A first and crucial 
difference between the DMRG and the PRG 
is that in the latter there are no distinction
between left and right blocks which get fused into a 
single one.

\subsection{ Wave-Function  Variational Ansatz}
\label{sec2:level2}

The superblock decomposition of the lattice in turn induces a
decomposition of the wave-function of the targeted states
associated  to the blocks and the sites. To be more precise,
let us assume that the total number of states kept during the
truncation/renormalization process is $N_E$, which includes the GS
and $N_E-1$ excited states. Then, the superblock wave-function
$\Psi_l({\bf n})$ at RG-step $l$ and lattice point ${\bf n}$ is
split into the following four pieces:

\begin{equation}
\Psi_l({\bf n}) =
\begin{cases}
\sum_{\alpha=1}^{N_E} a_{\alpha} L^{\alpha}_l({\bf n}) & \quad   {\bf n} \in B_l^L \\
a_{N_E+1}            & \quad   {\bf n} = \bullet_{l+1} \\
a_{N_E+2}            & \quad   {\bf n} = \bullet_{l+2} \\
\sum_{\alpha=1}^{N_E} a_{N_E+2+\alpha} R^{\alpha}_{l+3}({\bf n}) &
\quad {\bf n} \in B_{N-l-2}^R
\end{cases}
\label{3}
\end{equation}

\noindent where the symbols $\bullet_{l+1}$ and $\bullet_{l+2}$
denote two points separating  the blocks $B_l^L$ and $B_{N-l-2}^R$; 
$\{ L^{\alpha}_l({\bf n})\}_{\alpha=1}^{N_E}$ and
$\{ R^{\alpha}_{l+3}({\bf n})\}_{\alpha=1}^{N_E}$ are orthonormal
basis of states describing the degrees of freedom in the blocks
$B_l^L$ and $B_{N-l-2}^R$, respectively,i.e.,

\begin{equation}
\langle L^{\alpha}_l|L^{\beta}_l \rangle = \delta_{\alpha,\beta}
= \langle R^{\alpha}_{l+3}|R^{\beta}_{l+3} \rangle
\label{4}
\end{equation}

\noindent and the free unknown coefficients
$\{a_{\alpha}\}_{\alpha=1}^{2N_E+2}$ at this $l$-stage  will be
determined later on by means of a diagonalization/truncation
process defining the renormalization. They are normalized as

\begin{equation}
\parallel {\bf a} \parallel^2 = \sum_{\alpha=1}^{2N_E+2} a_{\alpha}^2 = 1
\label{5}
\end{equation}

Let us now present the PRG version of the SB wave-function.
To simplify matters we shall first consider the case
when $N_E=1$. Let $\Phi({\bf n})$ be a trial wave function 
describing the GS of the universe ${\cal U}$, with 
energy $E_\Phi = \langle \Phi |H | \Phi \rangle$. 
The first step of the method is to puncture the universe
at a given site, say $\bullet_l$, so that it can be seen
as the superblock $B_l \bullet$, where $B_l$ contains
all the sites of ${\cal U}$ but $\bullet_l$,
which is labelled by the integer $l$ denoting the RG step. 
To define the puncture operation we shall use
the projector operator onto the site $\bullet_l$, i.e.

\begin{equation}
P_{\bullet_l} | \Phi \rangle = \Phi(\bullet_l) | \bullet_l \rangle
\label{6}
\end{equation}

\noindent where  $| \bullet_l \rangle$ is the N-component vector
with 1 at the position corresponding to the site
$\bullet_l$. Using $P_{\bullet_l}$ the state $|\Phi \rangle$ decomposes as

\begin{equation}
|\Phi \rangle = N_l | M_l \rangle + \Phi(\bullet_l) |\bullet_l \rangle
\label{7}
\end{equation}

\noindent where $|M_l\rangle$ is the normalized vector

\begin{align}
& |M_l \rangle = \frac{1}{N_l} ( 1 - P_{\bullet_l}) |\Phi \rangle
\label{8} \\
& N_l = \sqrt{1 - \Phi(\bullet_l)^2 }
\nonumber 
\end{align}

\noindent which vanishes at the puncture
$\bullet_l$. The PRG ansatz consists in the following  
generalization of the state (\ref{7})

\begin{equation}
|\Psi \rangle = a_1 | M_l \rangle + a_2 |\bullet_l \rangle
\label{9}
\end{equation}

\noindent where $a_1, a_2$ are variational parameters
to be fixed later on  diagonalizing the SB Hamiltonian.
A trivial but important observation is that 
the state $|\Phi \rangle$ is in fact a particular case
of the state $|\Psi \rangle$ where $a_1 = N_l$ and 
$a_2 = \Phi(\bullet_l)$. 
The key idea behind the PRG method is the repairing
or improvement of $|\Phi \rangle$ at the site $\bullet_l$
by considering the dynamics of that site coupled to the
rest of the universe.

The puncture method can be inmediately 
generalized to the case with $N_E \geq 1$ states.
Let us denote by $\{\Phi^\alpha \}_{\alpha=1}^{N_E}$ a set of $N_E$
orthonormal states, i.e.

\begin{equation}
\langle \Phi^\alpha | \Phi^\beta \rangle = \delta_{\alpha, \beta}
\label{10}
\end{equation}

\noindent which give a variational approximation to the
 lowest lying states
of the Hamiltonian $H$. We shall also suppose that the restriction
of $H$ to the subspace expanded by this basis is diagonal, i.e.

\begin{equation}
\langle \Phi^\alpha |H| \Phi^\beta \rangle = \delta_{\alpha, \beta} 
\; E^\alpha_\Phi 
\label{11}
\end{equation}

\noindent where $E^\alpha_\Phi$ are the lowest $N_E$ energies
of $H$ at this stage of the RG. 
Projecting  out the site $\bullet_l$ from the basis 
 $\{\Phi^\alpha \}_{\alpha=1}^{N_E}$,  we obtain 
a set of $N_E$ vectors which are neither normalized
nor orthogonal. A new set of orthonormal vectors
$\{ M_l^\alpha \}_{\alpha=1}^{N_E}$ 
is obtained by means of the transformation

\begin{equation}
|M_l^\alpha \rangle = \sum_{\beta=1}^{N_E}
\; {\cal O}_{\alpha \beta} \; ( 1 - P_{\bullet_l}) \; |\Phi^\beta \rangle
\label{12}
\end{equation}

\noindent where the matrix ${\cal O}_{\alpha \beta}$ 
diagonalizes the scalar product of the punctured
states $( 1 - P_{\bullet_l}) |\Phi^\alpha \rangle$, namely,

\begin{equation}
\sum_{\alpha', \beta'}\; 
{\cal O}_{\alpha \alpha'} \; 
{\cal O}_{\beta \beta'}\; ( \delta_{\alpha', \beta'}
- \Phi^{\alpha'}(\bullet_l)   \Phi^{\beta'}(\bullet_l) )
= \delta_{\alpha, \beta}
\label{13}
\end{equation}

The analogue of eq.(\ref{7}) reads

\begin{equation}
|\Phi^\alpha \rangle =
\sum_\beta \; {\cal O}^{-1}_{\alpha \beta} \; 
|M^\beta_l \rangle +  \Phi^\alpha(\bullet_l) | \bullet_l \rangle
\label{14}
\end{equation}

\noindent where ${\cal O}^{-1}$ is the inverse of the
matrix ${\cal O}$. Eq.(\ref{14}) leads to the following
PRG ansatz

\begin{equation}
|\Psi \rangle =
\sum_{\alpha=1}^{N_E} \; a_\alpha  \; 
|M^\alpha_l \rangle +  a_{N_E +1}  | \bullet_l \rangle
\label{15}
\end{equation}

\noindent which for comparison with the DMRG eq.(\ref{3})
we write as,

\begin{equation}
\Psi_l({\bf n}) =
\begin{cases}
\sum_{\alpha=1}^{N_E} a_{\alpha} 
M^{\alpha}_l({\bf n}) & \quad   {\bf n} \in B_l \\
a_{N_E+1}            & \quad   {\bf n} = \bullet_{l} \\
\end{cases}
\label{16}
\end{equation}

\noindent The normalization condition on the $a$-parameters is:

\begin{equation}
\parallel {\bf a} \parallel^2 = \sum_{\alpha=1}^{N_E+1} a_{\alpha}^2 = 1
\label{17}
\end{equation}

\indent 
As in the $N_E =1$ case,  the wave functions $|\Phi^\alpha \rangle$
are a particular case of states $\Psi$ corresponding to the choices
$a^{(\alpha)}_\beta = {\cal O}_{\alpha \beta}^{-1}, \;\;
a^{(\alpha)}_{\bullet_l} = \Phi^\alpha(\bullet_l)$.
 The latter vectors are orthonormal and expand a hyperplane
of dimension $N_E$ embedded in the space of SB wave functions
which has dimension $N_E +1$. The PRG method for more than one puncture
is a straightforward generalization of the one puncture case  
 and we do not give the details.

Both in the DMRG and the PRG we are using the 
same symbol to denote the RG-parameters ${\bf a}$,
even though they come in different
number: a ($2N_E+2$)-dimensional 
vector for standard DMRG and a ($N_E+1$)-dimensional vector
for the PRG.

\subsection{  Superblock Hamiltonians}
\label{sec2:level2b}

The superblock Hamiltonians contain
the effective description of the lattice superblock
at a given step of the RG process. 
This is a reduced description of the whole Hamiltonian
for the original lattice. 
Thus, the dimensionality of these SB Hamiltonians is much smaller
than that of the original 
Hamiltonian and this makes their diagonalization something 
manageable. The dimension of $H_{SB}$ 
depends on the number of targeted states and 
the version of the RG 
method: $(2N_E+2)\times(2N_E+2)$ for the standard DMRG and 
$(N_E+1)\times(N_E+1)$ for the PRG.

\noindent In order to obtain 
the matrix elements of the superblock Hamiltonian
we need to identify the 
energy associated to the original Hamiltonian $H$
in the wave-function ansatzs 
of (\ref{3}) or (\ref{16}) with the reduced matrix elements
of the superblock Hamiltonian, namely, we demand

\begin{equation}
\langle  \Psi_l|H
|\Psi_l \rangle = \langle {\bf a}|H_{SB}(l)|{\bf a}\rangle 
\label{18}
\end{equation}

\noindent Inserting  
the respective ansatzs into  this relation one obtains an identity
between two quadratic 
forms in the variable ${\bf a}$, such that the input data is on
the LHS and the unknown matrix 
elements are on the RHS. To be more precise, for the standard
DMRG method the superblock 
Hamiltonian exhibits the following $4\times 4$ structure
associated to the four pieces of the DMRG decomposition of the lattice

\begin{equation}
H_{SB}(l) = 
\begin{pmatrix}
H_{L} & H_{L \bullet_{l+1}} &  H_{L \bullet_{l+2}} & H_{LR} \\
H_{\bullet_{l+1} L} & H_{\bullet_{l+1} 
\bullet_{l+1}} &  H_{\bullet_{l+1}\bullet_{l+2}} &
H_{\bullet_{l+1} R} \\
H_{\bullet_{l+2} L} & H_{\bullet_{l+2} 
\bullet_{l+1}} & H_{\bullet_{l+2}\bullet_{l+2}} &
H_{\bullet_{l+2} R} \\
H_{RL} & H_{R\bullet_{l+1}} & H_{R\bullet_{l+2}} & H_{R}
\end{pmatrix}
\label{19}
\end{equation}

\noindent Plugging the ansatz (\ref{3})   into 
(\ref{18}) we
arrive at the 
following expression for the various matrix elements in (\ref{19})

\begin{alignat}{2}
& H_{L}^{\alpha,\beta}  = \langle L_l^{\alpha}|H| L_l^{\beta} \rangle & \quad
& H_{R}^{\alpha,\beta}  = 
\langle R_{l+3}^{\alpha}|H| R_{l+3}^{\beta} \rangle \notag \\
& H_{\bullet_{l+1} \bullet_{l+1}}  
= \langle \bullet_{l+1} |H| \bullet_{l+1} \rangle & \quad 
& H_{\bullet_{l+2} \bullet_{l+2}}  
= \langle \bullet_{l+2}| H| \bullet_{l+2} \rangle \notag \\
& H_{LR}^{\alpha,\beta}  
= \langle L_l^{\alpha}|H| R_{l+3}^{\beta} \rangle & \quad 
& H_{\bullet_{l+1} \bullet_{l+2}}  
= \langle \bullet_{l+1} | H| \bullet_{l+2} \rangle \notag \\
& H_{L \bullet_{l+1}}^{\alpha} 
= \langle  L_l^{\alpha} |H | \bullet_{l+1} \rangle & \quad
& H_{L \bullet_{l+2}}^{\alpha} 
= \langle  L_l^{\alpha} |H | \bullet_{l+2} \rangle  \notag \\
& H_{R \bullet_{l+1}}^{\alpha} 
= \langle  R_{l+3}^{\alpha} |H | \bullet_{l+1} \rangle & \quad
& H_{R \bullet_{l+2}}^{\alpha} 
= \langle  R_{l+3}^{\alpha} |H | \bullet_{l+2} \rangle  
\label{20}
\end{alignat}

As it happens, all the 
matrix elements of the superblock Hamiltonian have a natural
geometrical meaning in 
terms of the interactions among the different parts of the
lattice superblock. 
For instance, those in the  
first line  describe the interactions only
within each of the left and right blocks; those in the second line 
represent the original 
interaction at the probing sites $\bullet_{l+1}$ and $\bullet_{l+2}$; 
those in the third line represent the interaction 
between the left and right blocks
and between the two probing sites, respectively; and so on and so forth.

\noindent Depending on the type 
of original interaction, the structure of the superblock
Hamiltonian will differ. 
As an example, for a problem with a local potential in real space,
the short-range structure 
is also transported onto the structure of the superblock. 
Thus, several matrix 
elements are vanishing, like $H_{LR}$, $H_{L\bullet_{l+2}}$ and
$H_{R\bullet_{l+1}}$ 
\cite{whitebook}, \cite{qm-dmrg}.
However, when the problem 
has long-range interactions then all matrix elements are
generically non-vanishing 
and the structure becomes more involved \cite{delta}.

%%%%%%%

Similarly, to obtain 
the superblock matrix elements in the PRG method we plug
ansatz (\ref{16}) into (\ref{18}) arriving  at the following expressions

\begin{equation}
H_{SB}(l) = 
\begin{pmatrix}
H_{B} & H_{B \bullet_{l}}  \\
H_{B \bullet_{l} } &   H_{\bullet_{l}}
\end{pmatrix}
\label{21}
\end{equation}

\begin{align}
& H_B^{\alpha, \beta}  = \langle M_l^\alpha|H|M_l^\beta\rangle \nonumber \\
& H_{B \bullet_{l}}^\alpha = \langle M_l^\alpha | H | \bullet_{l} \rangle 
\label{22} \\
& H_{\bullet_{l}}  = \langle \bullet_{l}| H|  
\bullet_{l} \rangle \nonumber  
\end{align}

\noindent Similar geometrical 
considerations apply to the meaning of the different
matrix elements in terms of interactions among block and punctures.

\noindent It is quite apparent 
that the superblock structure of the PRG method is much
simpler than the standard 
DMRG. This would imply in principle that the PRG is more economical
and advantageous than 
the DMRG. However, we shall see that there is a tradeoff between 
economy of resources and computational time during the sweeping process.

In the DMRG method the entries of the SB Hamiltonian
(\ref{19}) are data which are stored as functions of the
RG step $l$,  and they are updated after every step. 
However in the PRG method it is more
convenient to derive the SB Hamiltonian 
from more elementary data. 
Indeed let us consider again the case $N_E=1$, where
$H_B$ and $H_{B \bullet_l}$ are just two numbers.
Using eq.(\ref{8}) we find

\begin{align}
& H_B = \frac{1}{N_l^2}  \left( 
E_\Phi - 2 \Phi(\bullet_l) \langle \bullet_l |H |\Phi \rangle
+ \Phi^2(\bullet_l) \langle \bullet_l |H |\bullet_l \rangle
\right) 
\label{23} \\
& H_{B \bullet_{l}}= 
\frac{1}{N_l}  \left( 
\langle \bullet_l |H |\Phi \rangle
- \Phi(\bullet_l) \langle \bullet_l |H |\bullet_l \rangle
\right) \nonumber  
\end{align}

\noindent Hence we can construct the SB Hamiltonian
from the knowledge of $\Phi$, its energy $E_\Phi$
and the N-components
of $H|\Phi \rangle$, namely $\langle {\bf n}|H|\phi \rangle$.
Strictely speaking we only need  $\langle \bullet_l|H|\phi \rangle$
in (\ref{23}), but as one wants to move the puncture all over
the lattice one needs in practice to keep all the 
matrix elements  $\langle {\bf n}|H|\phi \rangle, \;\; 
\forall {\bf n}$. 

Recalling  that the state $\Phi$ is a special choice
of the general state $\Psi$, we derive the important
result that the lowest energy $E_0$ of the SB Hamiltonian
(\ref{21}) is  lower than  the energy $E_\Phi$
of the original ansatz, i.e.

\begin{equation}
E_0 \leq E_\Phi
\label{23b}
\end{equation}

\noindent which implies that the PRG flow always
lowers the energy of the ansatz.

\noindent 
When $N_E \geq 1$, using  eqs.(\ref{12}) and (\ref{22}) ,  we obtain
the entries $H_B, H_{B \bullet_l}$

\begin{align}
& H_B^{\alpha, \beta}  = 
\sum_{\alpha', \beta'} 
{\cal O}_{\alpha \alpha'} 
{\cal O}_{\beta \beta'}
( E_\Phi^{\alpha'} \delta_{\alpha' \beta'}
+ \Phi^{\alpha'}(\bullet_l) \Phi^{\beta'}(\bullet_l)
\langle \bullet_l | H| \bullet_l \rangle 
 \nonumber \\
& \;\;\;\,\,  - \Phi^{\alpha'}( \bullet_l) \langle
\bullet_l |H |  \Phi^{\beta'} \rangle 
 - \Phi^{\beta'}( \bullet_l) \langle
\bullet_l |H |  \Phi^{\alpha'} \rangle 
) \label{24} \\
& H_{B \bullet_{l}}^\alpha  = 
\sum_\beta {\cal O}_{\alpha \beta}
\; \left(  \langle
\bullet_l |H |  \Phi^{\beta} \rangle 
-  \Phi^{\beta}( \bullet_l) \langle
\bullet_l |H | \bullet_l \rangle  \right)
\nonumber 
\end{align}

\noindent in terms of $|\Phi^\alpha \rangle, H |\Phi^\alpha \rangle$ 
and $E^\alpha_\Phi$.

\subsection{ Truncation of Hilbert Space and Renormalization}
\label{sec2:level3}

The next step is to specify the projection of the wave functions onto
the several blocks and 
the free parameters ${\bf a}$ entering in the wave-function
ansatz (\ref{3}) and (\ref{16}), 
as well as the renormalization of the matrix elements
of the superblock Hamiltonian.

\noindent The starting point 
is the diagonalization of the superblock Hamiltonians
in order to obtain the 
wave functions of the $N_E$ targeted states corresponding to
the lowest energy eigenvalues. 
The free parameters ${\bf a}$ will be constructed 
out of the components of the targeted wave functions.

\noindent In the DMRG case, let us denote these parameters as the set
$\{ {\bf a}_L^{i}, a_{\bullet_{l+1}}^{i}, 
a_{\bullet_{l+2}}^{i}, {\bf a}_R^{i} \}_{i=1}^{N_E}$,
where ${\bf a}_L^{i}$ and ${\bf a}_R^{i}$ are $N_E$-dimensional
vectors.

The truncation of the Hilbert space is performed by the projection of the
superblock wave-functions onto the block formed by one block (left or right)
and the nearest probing site. Consequently, we have two possibilities to
perform such truncation, depending on whether we project onto the left block
or the right block, respectively, and the explicit form of these truncations
reads as follows:

\noindent a) $B_L \bullet \bullet B_R \rightarrow B_L \bullet$.

\begin{equation}
{\bf a}^{i} = 
\begin{pmatrix}
{\bf a}_L^{i} \\
a_{\bullet_{l+1}}^{i}\\
a_{\bullet_{l+2}}^{i}\\
{\bf a}_R^{i}\\
\end{pmatrix} \longrightarrow 
\begin{pmatrix}
{\bf a}_L^{i} \\
a_{\bullet_{l+1}}^{i}\\
\end{pmatrix}
\label{25}
\end{equation}

\noindent The projected wave functions in the RHS of (\ref{25}) must be
orthonormalized for they will become the new wave functions of the 
renormalized left block $B'_L$. Let us denote this process as

\begin{equation}
\begin{pmatrix}
{\bf a}_L^{\prime i} \\
a_{\bullet_{l+1}}^{\prime i}\\
\end{pmatrix} = \sum_j  {\cal O}^L_{i j} 
\begin{pmatrix}
{\bf a}_L^{j} \\
a_{\bullet_{l+1}}^{j}\\
\end{pmatrix}
\label{26}
\end{equation}

\noindent where ${\cal O}^L$ is the orthonormalization matrix
(using a Gram-Schdmit method, for instance).
If we were only targeting the GS wave function, then this would
amount to a simple normalization of the projected wave function,
namely, $a^{\prime}_L=a_L/N_L, a^{\prime}_{\bullet}=a_{\bullet}/N_L$
with $N_L=\sqrt{a_L^2 + a_{\bullet}^2}$.

\noindent The new wave functions for the renormalized left block 
$L_{l+1}^{\prime i}$
to be constructed in the next RG-step $l+1$ are given by

\begin{equation}
L_{l+1}^{\prime i}({\bf n}) = 
\begin{cases}
\sum_\alpha  a^{\prime i}_{L, \alpha} 
L_{l}^{\alpha}({\bf n}) & {\bf n}\in B^L_l \\
a^{\prime i}_{\bullet_{l+1}}                      & {\bf n}=\bullet_{l+1}\\
\end{cases}
\label{27}
\end{equation}

\noindent b) $B_L \bullet \bullet B_R \rightarrow \bullet B_R$.

\begin{equation}
{\bf a}^{i} = 
\begin{pmatrix}
{\bf a}_L^{i} \\
a_{\bullet_{l+1}}^{i}\\
a_{\bullet_{l+2}}^{i}\\
{\bf a}_R^{i}\\
\end{pmatrix} \longrightarrow 
\begin{pmatrix}
a_{\bullet_{l+2}}^{i}\\
{\bf a}_R^{i} \\
\end{pmatrix}
\label{28}
\end{equation}

\noindent Likewise, the projected wave functions in the RHS of (\ref{28}) 
must be orthonormalized for they will become the new wave functions of the 
renormalized left block $B_R$:

\begin{equation}
\begin{pmatrix}
a_{\bullet_{l+2}}^{\prime i}\\
{\bf a}_R^{\prime i} \\
\end{pmatrix} = \sum_j
{\cal O}^R_{i j} 
\begin{pmatrix}
a_{\bullet_{l+2}}^{j}\\
{\bf a}_R^{j} \\
\end{pmatrix}
\label{29}
\end{equation}

\noindent and similarly for the 
 new wave functions for the renormalized right block 
$R_{l+2}^{\prime i}$:

\begin{equation}
R_{l+2}^{\prime i}({\bf n}) = 
\begin{cases}
a^{\prime i}_{\bullet_{l+2}}                      & {\bf n}=\bullet_{l+2}\\
\sum_\alpha a^{\prime i}_{R, \alpha} R_{l+3}^{\alpha}({\bf n}) & {\bf n}\in B^R_{l+3} \\
\end{cases}
\label{30}
\end{equation}

Once the truncation process is carried out, we need to update the different
entries in the superblock Hamiltonian (\ref{19}). This is an asymmetric 
procedure, depending on whether we are renormalizing from left to right
as in a), or from right to left as in b). 
As an example we shall  give the renormalized left block Hamiltonian 
in case a). In matrix notation it reads,

\begin{equation}
H^{\prime}_L(l+1) = 
A^{\prime \dagger}
\begin{pmatrix}
H_{L} & H_{L \bullet_{l+1}}  \\
H_{L \bullet_{l+1}} & H_{\bullet_{l+1},\bullet_{l+1}}\\
\end{pmatrix}
A^{\prime}
\label{31}
\end{equation}

\begin{equation}
A^{\prime} = 
\begin{pmatrix}
{\bf a}_L^{\prime 1} & \ldots & {\bf a}_L^{\prime N_E} \\
a_{\bullet_{l+1}}^{\prime 1} & \ldots & a_{\bullet_{l+1}}^{\prime N_E}\\
\end{pmatrix}
\label{32}
\end{equation}

\noindent whereas the right block 
Hamiltonian $H^{\prime}_R$ is plainly taken from 
a previous step corresponding 
to the appropriate length, matching  the superblock.
These data yield  the superblock for the next RG-step,
namely, $B^{\prime L}_{l+1} \bullet \bullet B^{\prime R}_{N - l -3}$.

\noindent The case b) of right-to-left renormalization is similar 
and we skip the details.

Now, let us turn to the PRG method. As it happens, things are simpler to
formulate in this case: we do not need to distinguish between left-to-right
nor right-to-left cases anymore. There is only one way to truncate. First, we
diagonalize the superblock Hamiltonian (\ref{21}) and keep the  $N_E$ lowest
energy  eigenvectors out of $N_E+1$.

\noindent 
The case $N_E=1$ is specially simple since the SB Hamiltonian
is a $2 \times 2$ matrix which can be diagonalized
analytically. The GS energy and wave function are given by

\begin{align}
& E_0 = \frac{1}{2} \left( H_B + H_{\bullet_l} 
- \sqrt{ 4 \; H_{B \bullet_l}^2 + ( H_B - H_{\bullet_l})^2 } \right)
\nonumber \\
& {\hat a}_1 = \frac{1}{D} ,\;\;
{\hat a}_2 = \frac{ H_{B \bullet_l} }{ D ( E_0 - H_{\bullet_l})} \label{33} \\
& D = \sqrt{ 1 + \left( 
\frac{ H_{B \bullet_l} }{  E_0 - H_{\bullet_l}} \right)^2 }
\nonumber  
\end{align}

These eqs. are reminiscent of perturbation theory 
with  $H_B$ ( resp. $H_{B \bullet_l}$)  
playing  the role of the unperturbed ( resp. perturbed ) Hamiltonian.

The final step of the method is to replace the starting
wave function $\Phi$ by the state $\Psi$ given by eqs. (\ref{9})
and (\ref{33}), while the energy $E_\Phi$ is replaced
by the GS energy $E_0$, namely,

\begin{align}
& |\Phi \rangle  \rightarrow |\Phi' \rangle = 
|\Psi ({\hat a}_1, {\hat a}_2)  \rangle 
 \label{34} \\
& E_\Phi \rightarrow E_{\Phi'} = E_0 \nonumber
\end{align}

\noindent 
The old ( i.e. $\Phi$)  and new ( i.e.$\Phi'$)  
universe wave functions 
satisfy a simple relationship, which can be derived
from eqs.(\ref{8}) and (\ref{9}), namely

\begin{equation}
|\Phi' \rangle  = \frac{ \hat{a}_1}{ N_l}
\; |\Phi \rangle + \left( \hat{a}_2 - \frac{ \hat{a}_1 
\Phi(\bullet_l) }{ N_l} \right) \; |\bullet_l \rangle
\label{35}
\end{equation}

\noindent This eq. means that the wave function 
$\Phi'$ is a local perturbation of $\Phi$,
accompanied by an global rescaling 
outside the puncture. 
For example if  the value
of  $\Phi'$ at the puncture
is greater than in $\Phi$, i.e.
$| \Phi'(\bullet_l)| > |\Phi(\bullet_l)|$, 
then  outside the puncture
the new wave function will always be smaller
than in the original one, in order to preserve
the norm. This interpretation of the
PRG method will give us some hints to 
understand the numerical results 
presented in the next section.

\noindent From eq.(\ref{35}) we can also derive
the entries of $H |\Phi' \rangle $, which will be
needed for the next RG step,

\begin{equation}
\langle{\bf n} | H | \Phi' \rangle
= \frac{\hat{a}_1 }{ N_l}  
\langle{\bf n} | H | \Phi \rangle
+  \left( \hat{a}_2 - \frac{ \hat{a}_1 
\Phi(\bullet_l) }{ N_l} \right) \;
\langle {\bf n} | H  |\bullet_l \rangle
\label{36}
\end{equation}

\noindent Eqs.(\ref{34}), (\ref{35}) and (\ref{36}) 
completely define the renormalized state and we can 
move to a new puncture, say $\bullet_{l+1}$ to repeat
the process. 

\noindent 
For $N_E \geq 1$ we find the lowest $N_E$ 
states of the SB Hamiltonian. Denoting 
the corresponding wave functions as 
$({\bf \hat{a}}^i, \hat{a}^i_{\bullet_l})$ and the energies 
as $E^i$ with $i=1, \dots, N_E$, the analogue
of eqs.(\ref{35}) and (\ref{36}) read,

\begin{align}
& |\Phi^{\prime i} \rangle =
\sum_{\alpha, \beta} \;
\hat{a}^i_{\alpha} \; {\cal O}_{\alpha \beta}
\; | \Phi^\beta \rangle  + ( \hat{a}_{\bullet_l} -
\sum_{\alpha, \beta} \;
\hat{a}^i_{\alpha} \; {\cal O}_{\alpha \beta}
 \Phi^\beta(\bullet_l) )
 | \bullet_l \rangle \label{37}
\end{align}

\begin{align}
 \langle {\bf n} | H  |\Phi^{\prime i} \rangle & =
\sum_{\alpha, \beta} \;
\hat{a}^i_{\alpha} \; {\cal O}_{\alpha \beta}
\; \langle {\bf n} | H
 | \Phi^\beta \rangle   \label{38} \\
& +  ( \hat{a}_{\bullet_l} -
\sum_{\alpha, \beta} \;
\hat{a}^i_{\alpha} \; {\cal O}_{\alpha \beta}
\; \Phi^\beta(\bullet_l) ) \;   \langle {\bf n} | H
| \bullet_l \rangle \nonumber
\end{align}

%\begin{align}
%& |\Phi^{'i} \rangle = 
%\sum_{\alpha, \beta} \; 
%\hat{a}^i_{\alpha} \; {\cal O}_{\alpha \beta}
%\; | \Phi^\beta \rangle  + ( \hat{a}_{\bullet_l} - 
%\sum_{\alpha, \beta} \; 
%\hat{a}^i_{\alpha} \; {\cal O}_{\alpha \beta}
% \Phi^\beta(\bullet_l) )
% | \bullet_l \rangle \label{37}
%\end{align}

%\begin{align}
%& \langle {\bf n} | H  |\Phi^{'i} \rangle = 
%\sum_{\alpha, \beta} \; 
%\hat{a}^i_{\alpha} \; {\cal O}_{\alpha \beta}
%\; \langle {\bf n} | H
% | \Phi^\beta \rangle   \label{38} \\
%& +  ( \hat{a}_{\bullet_l} - 
%\sum_{\alpha, \beta} \; 
%\hat{a}^i_{\alpha} \; {\cal O}_{\alpha \beta} 
%\; \Phi^\beta(\bullet_l) ) \;   \langle {\bf n} | H
%| \bullet_l \rangle \nonumber 
%\end{align}

In summary, the PRG method
consists in the following  steps:  puncture 
the universe at a given site, study 
the block-puncture dynamics and sewing
of the puncture on  the block,  getting a 
new effective description of the universe, i.e.

\begin{equation}
{\cal U} \overset{ \text{ puncture} }{\longrightarrow}
B \; \bullet_l 
 \overset{ \text{ sew } }{\longrightarrow}
{\cal U'}
\label{39}
\end{equation}

\subsection{ The Sweeping Process}
\label{sec2:level4}

So far we have defined the RG process at a given step for 
both the standard
DMRG an the PRG methods.
We have then the basic ingredients and we must say how the process carries on.
Namely, this is the sweeping process which amounts to specify how the original
lattice is traversed by moving the probing sites and punctures, respectively,
throughout the lattice.

\begin{figure}
\centering
\includegraphics[width=8 cm]{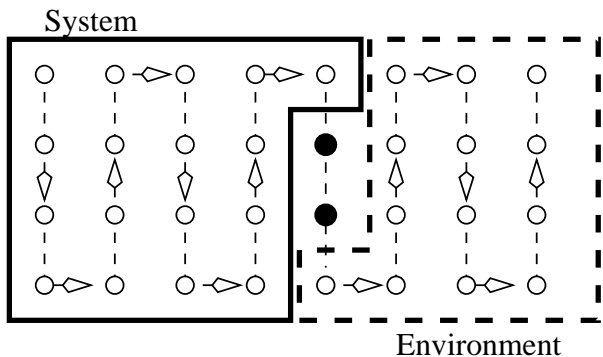}% Here is how to import EPS art
\caption{Schematic representation of the sweeping process in the 
standard DMRG method. The two black circles represent the probing sites
which move throughout the 2D lattice onto the embedded dashed line 
which serves as a guide.}
\label{fig3}
\end{figure}

In the DMRG case, the sweeping process is graphically described 
in Fig.~\ref{fig3}. 
The probing sites $\bullet \bullet$ move  from left to right,
which enlarges the left block upon renormalization, i.e. 

\begin{equation}
B_l^L \bullet_{l+1} \longrightarrow B_{l+1}^{\prime L}
\label{40}
\end{equation}

\noindent and similarly from right to left.

%%%%%%%%%%%%%%%%%%%%%%%%%%%%%%%%%%%%%%%%%%%%%%%%%%%%%%%%%%%%%%%%%%%%%%%%%%

%%%%%%%%%%%%%%%%%%%%%%%%%%%%%%%%%%%%%%%%%%%%%%%%%%%%%%%%%%%%%%%%%%%%%%%%%%%%%

\noindent In two dimensions, the sweeping process is acomplished by embedding
a one dimensional path  into the whole lattice.
This line is where the probing sites move along  as the RG process proceeds.

\begin{figure}[H]
\begin{center}
\includegraphics[width=4 cm, height=14 cm]{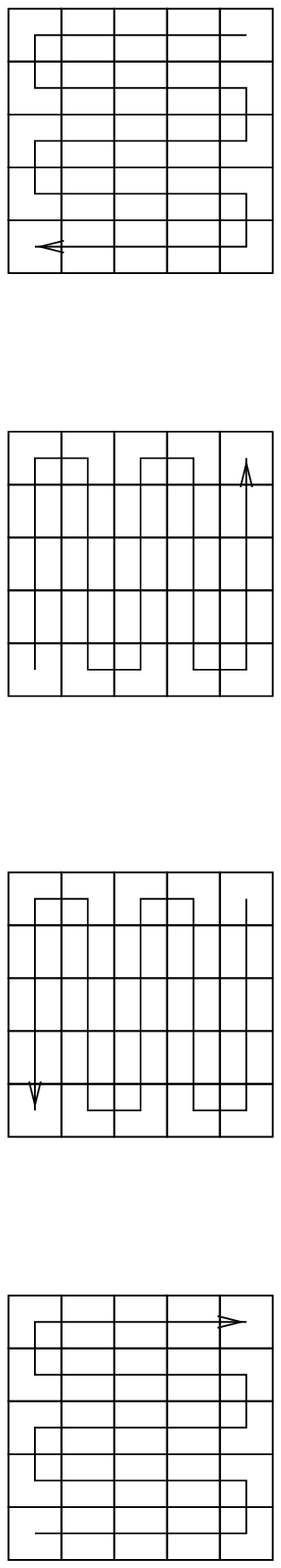}% Here is how to import EPS art
\end{center}
\caption{Schematic representation of the sweeping process in the 
PRG method. Here the center of the plaquettes represent the actual
sites of the real 2D lattice. Starting from the upper rightmost site,
the lines indicate the successive steps executed to complete a whole 
cycle or sweeping.}
\label{fig4}
\end{figure}

In the PRG method the puncture can move  through the lattice
following several patterns. 
The most convenient 
is when each site is traversed 
4 times in a cycle or sweep: left to right,
right to left, up to down and down to up. This is graphically depicted in 
Fig.~\ref{fig4}. 
\noindent Every  sweep  starts at the upper rightmost site and continues
with a left movement towards the lower leftmost site of the lattice, following
a zigzag line similar to that of the DMRG sweeping. At this moment
the puncture comes back to the original site
but following a down-to-up zigzag path. So far, each lattice site has been 
visited twice. 
To complete the sweeping cycle, there is another movement, this time
with a up-to-down zigzag path, towards the lower leftmost site. Finally,
the puncture comes back to the initial site with a zigzag down-to-up path.

%%%%%%%%%%%%%%%%%%%%%%%%%%%%%%%%%%%%%%%%%%%%%%%%%%%%%%%%%%%%%%%%%%%%%%%%%%%%%

%%%%%%%%%%%%%%%%%%%%%%%%%%%%%%%%%%%%%%%%%%%%%%%%%%%%%%%%%%%%%%%%%%%%%%%%%%%%%

%%%%%%%%%%%%%%%%%%%%%%%%%%%%%%%%%%%%%%%%%%%%%%%%%%%%%%%%%%%%%%%%%%%%%%%%%%%%%%%
Prior to the sweeping process, there is  a warmup process
to build up the finite lattice.
There is no restriction about the initial data of the
wave functions, but it is obvious that a smart guess shall boost the
process. In the 1D and 2D cases we have prepared such a guess using a
``Kadanoff-like'' warmup: the original lattice is coarse-grained until
the number of degrees of freedom is reduced to a given (small) value. Then
the effective hamiltonian at this scale is exactly diagonalized and the
resulting eigenfunctions are extrapolated to cover the whole original
lattice.

In the 3D case we choose the initial data to be a gaussian random vector
without correlation between different cells. Although it takes longer to
achieve the desired precision, convergence is also completely fulfilled.
%%%%%%%%%%%%%%%%%%%%%%%%%%%%%%%%%%%%%%%%%%%%%%%%%%%%%%%%%%%%%%%%%%%%%%%%%%

%\newpage
\section{Results in Several Dimensions}
\label{sec3:level1}

In this section we present our numerical results. 
We have made an extensive
comparative analysis between the DMRG and the
PRG involving different Hamiltonians, 
lattice's size and dimensionality,
number of states kept and number of punctures.  
We consider 
 short-range and long-range
interactions, in order 
to see whether the performance of the RG methods depends on this
fact. We present our results in the following subsections according to the
dimensionality of the lattices.

\subsection{One-Dimensional Models}
\label{sec3:level2}

We have solved 3 types of potentials with the two RG methods:

\noindent {\em Particle-in-a-box Potential}: 
this is the free particle case, i.e. 
$V_n=0, \forall n$.  The eigenfunctions are delocalized 
in real space.

\noindent {\em Harmonic Oscillator:} 
this is the standard quadratic potential $V_n=n^2$.
The eigenfunctions are 
localized in the centre of the lattice.

\noindent {\em 2D Delta Potential}: 
this model is  a discretization, in momentum
space,  of the two-dimensional 
delta function potential \cite{delta}. 
Due to the rotational symmetry
the Hamiltonian is 
 one-dimensional and it is given by

\begin{equation}
H_{n m} = \delta_{n,m} b^{2 n} - g \; b^{n+m}, \;\; M \leq n , m \leq N
\label{41}
\end{equation}

\noindent where $n,m$ are momentum labels, 
$M$ and $N$ play the role of infrared and 
ultraviolet cutoffs, respectively, 
$b$ is a control parameter of the discretization
which here we take as 
$b=\sqrt{2}$ and $g$ is the coupling constant of the delta function
\cite{delta-g}.

\begin{table}[h]
\begin{ruledtabular}
\begin{tabular}{lcddcdd}
\multicolumn{1}{c}{METHOD}  
  &\multicolumn{2}{c}{PBOX\footnotemark[1]}
  &\multicolumn{2}{c}{OSC\footnotemark[2]}
  &\multicolumn{2}{c}{DELTA\footnotemark[3]}
\\
  &\multicolumn{1}{c}{Sweeps}
  &\multicolumn{1}{c}{Time\footnotemark[4]}
  &\multicolumn{1}{c}{Sweeps}
  &\multicolumn{1}{c}{Time\footnotemark[4]}
  &\multicolumn{1}{c}{Sweeps}
  &\multicolumn{1}{c}{Time\footnotemark[4]}
\\
\colrule
Exact &          & 0.27 &        &  0.26 &       & 0.02 \\
DMRG  &           2 & 0.36      & 2 & 0.27       & 4 & 1.5 \footnotemark[5]\\
PRG (2) &         65 & 13.92    & 37 & 7.69      & 27  & 3.09 \\
PRG (4) &        48  & 4.75     & 25 & 2.36      & 10  & 1.25 \\
PRG (10)&        34  & 2.31     & 19 & 1.25      & 4   & 0.65 \\
\end{tabular}
\end{ruledtabular}
\footnotetext[1]{Particle-in-a-box potential.}
\footnotetext[2]{Harmonic oscillator.}
\footnotetext[3]{2D Delta potential.}
\footnotetext[4]{CPU seconds in a Pentium III  at 450 MHZ.}
\footnotetext[5]{Here the precision is $10^{-4}$.}
\label{table6a}
\caption{DMRG versus PRG numerical 
results for 1D systems. For all  models $N_E=4$
and the number of sites is $N=100$ for PBOX and OSC, while $N=38$ for DELTA.
The required precision is $10^{-10}$ for the GS and the first excited states.
In the PRG method the number 
in parentheses represents the number of punctures $N_p$.
}
\end{table}

In Table I we present a summary 
of our numerical results.
We have choosen the number 
of sweeps as the basic quantity to establish the comparison
between the two RG methods. 
This is because this number only depends on the RG method
employed for each model and not on the computer machine. 
However, to get an idea of what this 
number means we also give the corresponding computer time.
Moreover, as it happens the time spent in a DMRG sweep is different than
in a PRG sweep because the renormalization operations are also
different. 
We have also solved these models using  exact
diagonalization techniques.

\noindent 
Table I shows  that the number of sweeps needed by the DMRG to
achieve the prescribed 
convergence ($10^{-10}$ for the $N_E/2$ lowest energy states)
is much smaller than in the PRG method. This 
fact is independent of the 
model, the lattice size, 
the number of punctures $N_p$ 
and the  number of targeted states $N_E$.
We also observe that increasing the number of punctures
lowers the number of PRG  sweeps, which is nevertheless
greater than the DMRG ones.
On the other hand it is not reasonable 
to use  a large  number of punctures $N_p$, as compared to the total
number of sites,  
for this amounts to an almost 
exact diagonalization of the model.

The bad convergence of the PRG method in 1D is due to the
``rigidity'' of the ansatz,  meaning that the  updates
of the wave function affects  in the same amount its
left and right handed pieces. The high performance of the
DMRG can be attributed  to the ``flexibility'' of the
ansatz,  where the left and right handed pieces
of the ansatz are updated independently.

\subsection{Two-Dimensional Models}
\label{sec4:level2}

In two dimensions 
we have solved two models: the free particle and  the Hydrogen atom
whose  Coulomb potential,  
in atomic units,  takes the following form

\begin{equation}
V_{{\bf n}} = -\frac{2Z}{\sqrt{n_1^2+n_2^2}}
\label{42}
\end{equation}

\noindent with $Z$ the 
atomic number which is fixed to 1. 
This problem is factorizable and  
admits an analytical solution 
in the continuum, with energies given by $E_n=-4Z^2/n^2$.
We adopt a lattice discretization of (\ref{42})
where the origin is placed at the centre of a plaquette.

%new
%%%%%%%%%%%%%%%%%%%%%%%%%%%%%%%%%%%%%%%%%%%%%%%%%%%%%%%%%%%%%%%%%%%%%%%%%%%%%

\begin{table}[h]
\begin{ruledtabular}
\begin{tabular}{lcddc}
\multicolumn{1}{c}{METHOD}
  &\multicolumn{2}{c}{2D-HYDROGEN\footnotemark[1]}
  &\multicolumn{2}{c}{2D-PBOX\footnotemark[2]}
\\
  &\multicolumn{1}{c}{Sweeps}
  &\multicolumn{1}{c}{Time\footnotemark[3]}
  &\multicolumn{1}{c}{Sweeps}
  &\multicolumn{1}{c}{Time\footnotemark[3]}
\\
\colrule
%DMRG  &           2 & 0.36      & 2 & 0.27       \\
DMRG  &        9 & 45.6  & 12 & 65.7 \\
PRG (1$\times$1) &  15 & 5.85 & 15 & 5.85         \\
PRG (2$\times$2) &   5 & 3.23 &  5 & 3.22      \\
PRG (3$\times$3)&  3 & 3.33  &  3 & 3.32    \\
\end{tabular}
\end{ruledtabular}
\footnotetext[1]{2D Hydrogen atom}
\footnotetext[2]{Particle-in-a-box potential in 2D.}
\footnotetext[3]{CPU seconds in a Pentium III  at 450 MHZ.}
%\footnotetext[5]{Here the precision is $10^{-4}$.}
\label{table2}
\caption{DMRG versus PRG numerical results for 2D systems.
For all  models $N_E=4$
and the number of sites is $N=10\times 10$
The required precision is $10^{-10}$ for the GS and first excited state.
In the PRG method the number in parentheses represents the number
of punctures $N_p$ (see Fig.~\ref{fig2}).
}
\end{table}

%%%%%%%%%%%%%%%%%%%%%%%%%%%%%%%%%%%%%%%%%%%%%%%%%%%%%%%%%%%%%%%%%%%%%%%%%%%%% 

In table II we display  results for a $10 \times 10$
lattice, which show that
the PRG is more efficient than the DMRG.
This  fact is independent on the size
of the lattice, the number of states kept
and the model.  The aforementioned ``rigidity''
of the PRG ansatz seems to be quite appropiate
for 2D systems, while the left-right structure 
of the DMRG becomes a handicap. 
 
The $N_p=4$ PRG  reduces considerably
the number of sweeps while the consumed (i.e. CPU) 
 time is almost the same
as for $N_p=9$. Both the number of sweeps and the CPU
time for the PRG are quite insensitive to the model.

%%%%%%%%%%%%%%%%%%%%%%%%%%%%%%%%%%%%%%%%%%%%%%%%%%%%%%%%%%%%%%%%%%%%%%%%%%

\begin{figure}%[H]
\centering
\begin{center}
\includegraphics[height=7 cm, width= 7.5 cm]{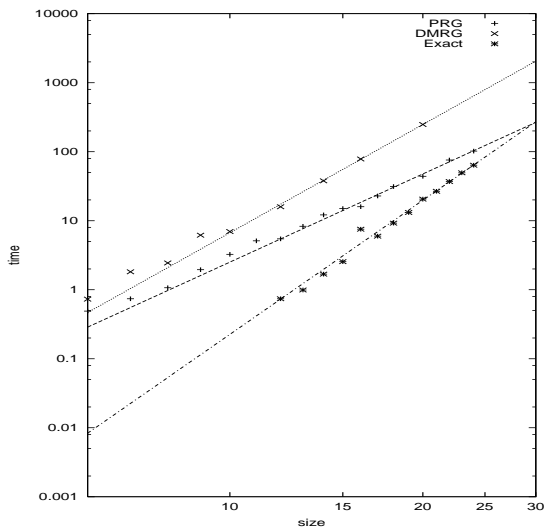}% Here is how to import EPS art
\caption{ 
Log--log plot, time $t_{{\rm CPU}}$ versus length size $L$,  for  the 2D
Hydrogen spectrum on a square lattice. Four states are targeted in 
each case, and a relative precision of $10^{-10}$ 
is required on the ground state and  first excited state.
The PRG uses a 2$\times$2 puncture as in Fig. 2.}
\label{fig5}
\end{center}
\end{figure}

%%%%%%%%%%%%%%%%%%%%%%%%%%%%%%%%%%%%%%%%%%%%%%%%%%%%%%%%%%%%%%%%%%%%%%%%%%%%%

In fig.~\ref{fig5}
we plot the CPU time $t_{{\rm CPU}}$ 
versus the linear size of  the square lattice $N_1=N_2\equiv L$
for the H-atom ( for the free particle we get similar results).  
From this plot we derive the scaling
law 

\begin{equation}
t_{{\rm CPU}} \propto L^{\theta}
\label{43}
\end{equation}

\noindent Table III collects  the numerical
values of the exponent $\theta$  
for the Exact, DMRG and PRG methods. 
They  are the slopes
of the curves in Fig.~\ref{fig5}.

%%%%%%%%%%%%%%%%%%%%%%%%%%%%%%%%%%%%%%%%%%%%%%%%%%%%%%%%%%%%%%%%%%%%%%%%%%%%%%%%%%%%%%%%%%%%%%%%
\begin{table}[h]
\begin{ruledtabular}
\begin{tabular}{lcc}
METHOD    &    PBOX     &   2D-HYDROGEN \\
\colrule
Exact     & $6.4\pm 0.3$  & $6.61\pm 0.27$\\
DMRG      & $5.5\pm 0.1$  & $5.21\pm 0.05$\\
PRG       & $3.92\pm 0.1$ & $3.85\pm 0.07$\\
\end{tabular}
\end{ruledtabular}
\caption{Values of the exponent $\theta$ in 
(\ref{43}) obtained with a linear fit
of the data plotted in Fig.~\ref{fig5}.
\label{table3}}
\end{table}
%%%%%%%%%%%%%%%%%%%%%%%%%%%%%%%%%%%%%%%%%%%%%%%%%%%%%%%%%%%%%%%%%%%%%%%%%%%%%%%%%%%%%%%%%%%%%

\noindent Fig.~\ref{fig5}  
shows that the elapsed time 
is always larger  for the DMRG than
for the PRG. Moreover, 
the $\theta$ exponent in the law (\ref{43}) is also larger in
the DMRG. This  confirms  
that in 2D the PRG method is more efficient than the 
standard DMRG. Table III shows that this conclusion remains true for both
types of models. For sufficiently large lattices the PRG performs
better than the exact diagonalization methods.

As an illustration of how the PRG method works during the renormalization
process leading to the determination of the lowest lying states of the
models, we present in Figs. 7   and 8
a set of typical  snapshots  representing
the four targeted wave functions in a 2D Hydrogen model and free particle. 
We do not plot the three dimensional picture of these wave functions, but 
instead the figure shows their projection onto the x-y plane, so that 
each grey square plaquette is more intense the higher the height (in absolute
value) of the wave function. The punctures of the PRG method
($2 \times 2$ and $1 \times 1$) appear in these figures as 
blanck and  white plaquettes respectively.

\noindent Making  
the superposition of all the screenshots of this sort,
one for each of the RG-steps, one  
produces a movie showing the 
convergence process of the targeted wave functions starting from the 
warm-up initial states. During this ``time'' evolution one sees how the wave
functions get shaped towards their final exact forms.

\subsection{Three-Dimensional Models}
\label{sec5:level1}

We believe that the computations 
presented in one- and two-dimensional lattices
are enough to make a 
thorough comparison of the two RG methods. However, we have
also carried out calculations 
in three-dimensional lattices. The main purpose of
this extension is to test the scaling laws obtained in the previous subsection
for the PRG method. As far as the PRG version is concerned, there is no need
for an extra effort in the computation and the sweeping process is a repetition
of the one employed in two-dimensions, but traversing all the parallel planes
making up the whole 3D lattice.
However, we have not done the similar extension for the DMRG method for it was
apparent from the previous 
subsection that the higher the dimensionality, the more
unnatural  the standard DMRG 
sweeping becomes. Thus, we expect a worse behaviour
in the DMRG method as the dimension increases.

%%%%%%%%%%%%%%%%%%%%%%%%%%%%%%%%%%%%%%%%%%%%%%%%%%%%%%%%%%%%%%%%%%%%%%%%%%%%%%%
\begin{figure}[H]
\centering
\includegraphics[height=7 cm, width= 7.5 cm]{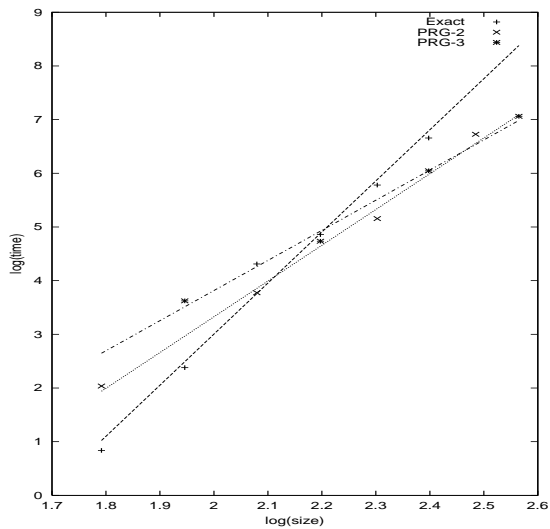}% Here is how to import EPS art
\caption{ Same plot as Fig.~\ref{fig5}, but for 3D Hydrogen atom. PRG is applied with a
$2\times 2$ and a $3\times 3$ punctures. Notice the crossing
between PRG  lines and the exact ones.}
\label{fig6}
\end{figure}

%%%%%%%%%%%%%%%%%%%%%%%%%%%%%%%%%%%%%%%%%%%%%%%%%%%%%%%%%%%%%%%%%%%%%%%%%%%%%

We have chosen the three-dimensional
Hydrogen atom as the representative model and thus, we have discretized the 
Coulomb potential

\begin{equation}
V_{{\bf n}} = -\frac{2Z}{\sqrt{n_1^2+n_2^2+n_3^2}}
\label{44}
\end{equation}

\noindent with the proviso that 
the origin is placed outside the lattice, at the 
center of a cube, to avoid the singularity of the potential.

%%%%%%%%%%%%%%%%%%%%%%%%%%%%%%%%%%%%%%%%%%%%%%%%%%%%%%%%%%%%%%%%%%%%%%%%%%%%%%%%%%
\begin{table}[h]
\begin{ruledtabular}
\begin{tabular}{lc}
METHOD    &      3D-HYDROGEN \\
\colrule
Exact &  $9.51 \pm 0.6$\\
PRG 2$\times$2 & $6.6 \pm 0.4$\\
PRG 3$\times$3 & $5.6 \pm 0.3$\\
\end{tabular}
\end{ruledtabular}
\caption{Values of the exponent $\theta$ in (\ref{43}) 
obtained with a linear fit
of the data plotted in Fig.~\ref{fig8}.
\label{table4}}
\end{table}
%%%%%%%%%%%%%%%%%%%%%%%%%%%%%%%%%%%%%%%%%%%%%%%%%%%%%%%%%%%%%%%%%%%%%%%%%%

Fig.~\ref{fig6} and table IV 
show that the $\theta$ exponents are larger than
those in 2D. However, the 3D exponent 
for PRG is again lower than the corresponding one  for the
exact diagonalization method. In fact, the PRG exponent 
is not much larger than the DMRG exponent for
2D lattices. This fact makes us to believe that 
the PRG version is a well-behaved procedure in dimensions
than one.
Moreover, the fact that increasing 
the number of punctures reduces the number of sweeps is also reflected in 
this table by the smaller values of the $\theta$ exponent.

%%%%%%%%%%%%%%%%%%%%%%%%%%%%%%%%%%%%%%%%%%%%%%%%%%%%%%%%%%%%%%%%%%%%%%%%%%%%%

\begin{figure*}
\includegraphics[width=15 cm]{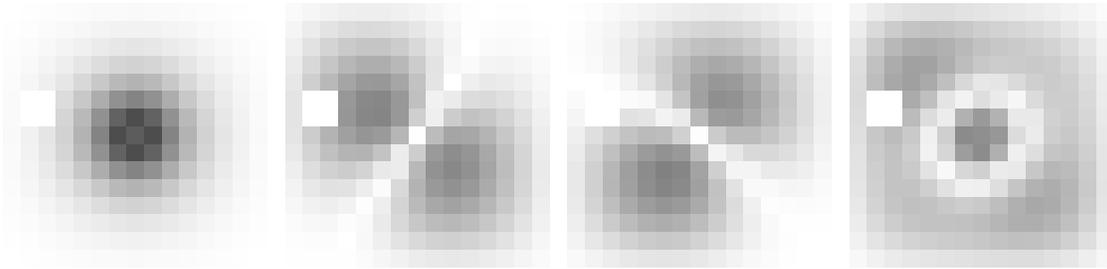}% Here is how to import EPS art
\caption{A typical screenshot of the PRG $2\times 2$ procedure for the
2D hydrogen atom on a $15\times 15$ square lattice and four target states.
Notice the puncture near the upper--left corner. The puncture moves through
the picture in a sweeping path, as depicted inf Fig.~\ref{fig4}.}
\label{fig7}
\end{figure*}
%%%%%%%%%%%%%%%%%%%%%%%%%%%%%%%%%%%%%%%%%%%%%%%%%%%%%%%%%%%%%%%%%%%%%%%%%%%%%

%%%%%%%%%%%%%%%%%%%%%%%%%%%%%%%%%%%%%%%%%%%%%%%%%%%%%%%%%%%%%%%%%%%%%%%%%%%%%

\begin{figure*}
\includegraphics[width=15 cm]{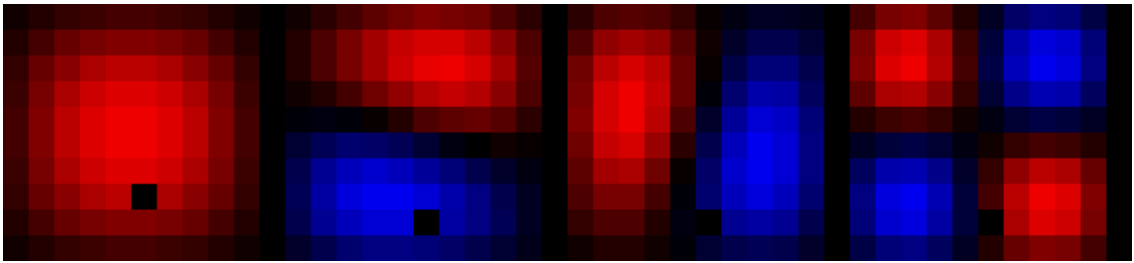}
\caption{A typical screenshot of the PRG $1\times 1$ procedure for the
2D free particle  on a $10\times 10$ square lattice and four target states.
Notice the puncture as a black plaquette. The puncture moves through
the picture in a sewing path, as depicted inf Fig.~\ref{fig4}.}
\label{fig8}
\end{figure*}
%%%%%%%%%%%%%%%%%%%%%%%%%%%%%%%%%%%%%%%%%%%%%%%%%%%%%%%%%%%%%%%%%%%%%%%%%%%%%

%\newpage
\section{Conclusions and Prospects}
\label{sec4:level1}

In this paper we have explored the formulation of a single-block
version of the DMRG method. The main motivation to undertake this
study is the construction of a version of the DMRG which is better
suited for higher dimensional lattices than the standard version.

\noindent We have stressed  the role played by the
sweeping process in the DMRG method  and we have singled it out as one of the
most relevant  features of the finite-system formulation of the DMRG.
It is this sweeping process  what becomes one of the defining
components of the PRG version of the DMRG and it is responsible
for achieving the convergence of the lowest energy properties 
to a certain prescribed precision.

\noindent In 1D lattices the standard DMRG method
outperforms the new PRG version. This is natural for we know that the
DMRG is somehow optimal when dealing with one-dimensional models.

\noindent However, in 2D lattices the PRG formulation is
more natural and well adapted to this type of lattices, unlike the DMRG
which needs to split the lattice into left and right blocks.
We have also tested numerically that the PRG gives a better
performance than the DMRG for several types of models.

\noindent We have checked that the PRG method also perfoms well
for 3D lattices. As a byproduct of this work, the PRG version
can be considered as a new  numerical method for solving the
Schr\"odinger equation in 2D and 3D with a better efficiency 
than the exact diagonalization techniques, as can be seen
from Tables III and IV.

Altogether we find these results quite promising,  
but of course the crucial issue  is wether one
can generalize the PRG to interacting many body
systems. The technical  point is to define
a ``puncture operation'' of many body wave functions,
which must isolate the ``local'' states associated with 
the puncture from the ``global'' states  
 associated with the block. As we have shown 
in quantum mechanics
this can be achieved by a set of local projection
operators which ``pick up'' the value of the wave
function at any given site ( see eq. \ref{6}). 
In the many body case one needs 
projection operators for all possible local
states of the puncture. 
The formalism that is  required to 
generalize the PRG to many body systems is
reminiscent to the Matrix Product (MP) method  
\cite{progress}.  
This fact may not be that surprising since after all the
variational state underlying the DMRG is an inhomogenous  
MP ansatz.

\begin{acknowledgments}

We would like to thank T. Nishino, R. Noack and  S.R. White 
for conversations.  J.R.-L. thanks S.N.~Santalla for some technical support. 
This work has been partially supported by the Spanish grant PB98-0685.

\end{acknowledgments}


\begin{thebibliography}{999}
\bibitem{delta-g} The coupling constant is set to 
$g= 0.06060600032108866$. This value
of $g$ is choosen in such a way 
that the exact ground state energy of (\ref{41}) 
is given exactly by $E_0=-1$ \cite{delta}.
\end{thebibliography}
\end{document}